\documentstyle[aps,multicol,epsf]{revtex}

\begin{document}
\title{
SU(4) spin-orbit critical state in one dimension
}
\author{
Yasufumi Yamashita, Naokazu Shibata,\cite{address} and Kazuo Ueda
}
\address{
Institute for Solid State Physics, University of Tokyo, 
Roppongi 7-22-1, Minatoku, Tokyo 106-8666, Japan
}
\date{\today}
\maketitle

\begin{abstract}
Effect of quantum fluctuations concerned with the orbital degrees of
freedom is discussed for the model with SU(4) symmetry in one dimension.
An effective Hamiltonian is derived from the orbitally degenerate Hubbard
model at quarter filling.  This model is equivalent to the Bethe soluble
SU(4) exchange model. Quantum numbers of the ground state and the 
lowest branch of excitations are determined. The spin-spin correlation
functions are obtained numerically by the density matrix renormalization
group method.  It shows a power-law decay with oscillations of 
the period of four sites.
The period originates from the interference between the spin and 
orbital degrees of freedom. The exponent of the power-law decay
estimated from the finite size data is consistent with the 
prediction by the conformal field theory. 
\end{abstract}
\begin{multicols}{2}
\narrowtext

\newpage\

\vspace*{-3ex}
\section{Introduction}               
Recently the role of the orbital degrees of 
freedom in strongly correlated electron systems 
is attracting growing interest. 
The increase of this attention is stimulated by  
the progress in the experimental studies of 
transition metal and rare earth compounds such 
as ${\rm{LaMnO_{3}}},{\rm{CeB_{6}}}$, 
and ${\rm{TmTe}}$, which show various 
interesting properties associated with the orbital degrees of freedom. 

In the 1970's Kugel and Kohmskii\cite{kugel} and Inagaki\cite{inagaki} 
studied an orbitally degenerate model 
to understand the magnetic structures of transition metal compounds 
within the mean field theory. 
They concluded that if orbitals ordered 
antiferromagnetically, then spins ordered 
ferromagnetically, and vice versa. 
Recently Shiina, Shiba, and Thalmeier\cite{shiinaC}
have studied similar models in connection with a quadrupolar 
ordering of CeB${_6}$ and discussed the phase diagram under 
an external magnetic field neglecting quantum fluctuations. 

In the case of ${\rm{LaMnO_{3}}}$, 
the orbital ordering temperature, ${{T_{O}}}$ ($\sim$ 775 K), 
is much higher than the 
N\'{e}el temperature, ${{T_{N}}}$ ($\sim$ 141 K), so the
mean field theoretical approaches are considered to be a good
starting point. 
On the other hand, for ${\rm{CeB_{6}}}$, ${{T_{O}}}$ 
($\sim$ 3.4 K) is the  
same order as ${{T_{N}}}$ ($\sim$ 2.3 K) and thus
the interplay between spin and orbital quantum fluctuations 
may 
be important. Therefore it is necessary to consider the effects 
of quantum fluctuations more seriously beyond the mean field theory. 

Before considering the effects of the orbital degrees of freedom, 
we briefly summarize the properties of 
the one-dimensional single orbital Hubbard model for comparison. 
In the limit of strong correlation at half filling, 
the model is reduced to the spin-1/2 antiferromagnetic (AF) 
Heisenberg model with SU(2) symmetry. 
This model is well known as a typical quantum 
critical system. 
The ground state of this model is singlet 
($[1^{2}]$ in Young's diagram representation) 
and the elementary excitations are 
gapless and triplet ($[2^{1}]$), so-called 
des Cloizeaux-Pearson modes\cite{pearson}. 
These results are consistent with the 
Lieb-Schultz-Mattis theorem,\cite{lieb,affleck} 
which states that the 
half-integer-$S$ spin chain, which has the translational 
and rotational symmetries, either has a singlet ground state with 
gapless excitations or has a finite gap with degenerate 
ground states, corresponding to spontaneous breaking of the parity. 

In the present paper we study an effective model of an orbitally 
degenerate Hubbard model in one dimension.
By the density matrix renormalization group (DMRG) 
method and exact diagonalization (Lanczos method),
we find a quantum critical state at the SU(4) symmetric
point, which originates from the strong interplay 
between spin and orbital 
quantum fluctuations in one dimension.

\section{Model}
\begin{center}
\begin{figure}
\leavevmode
\epsfxsize=86mm
\epsffile{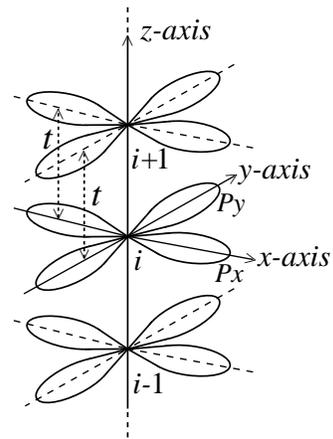}
\vspace*{0mm}
\caption{ A one-dimensional Hubbard model 
with twofold degenerate orbitals at each site.
}\label{model_fig}
\end{figure}
\end{center}

We start from the one-dimensional orbitally twofold degenerate 
Hubbard model with Hund rule coupling between the two orbitals at the 
same site. This is the simplest model which possesses 
orbital degrees of freedom. Hamiltonian of this model is given by
\begin{eqnarray}
H&=&{H_{t}}'+H_{U}+H_{J} \nonumber \\
 &=&\sum_{i \alpha \alpha ' \sigma }
\left(-t^{\alpha \alpha '}_{i,i+1} c^{\dagger}_{i\alpha \sigma}
c_{i+1\; \alpha ' \sigma}+{\rm{H.c}}\right) \nonumber \\
 &&+
\frac{U}{2}\sum_{i\alpha \alpha '\sigma \sigma '}
\Big{\{} n_{i \alpha \sigma}n_{i \alpha ' \sigma '}
\left( 1-\delta _{\alpha\alpha '}  \delta _{\sigma\sigma '}  
\right) \Big{\}} \nonumber \\
 &&-J\sum_{i}
\left(2{\vec{S}}_{i1}\cdot {\vec{S}}_{i2}+\frac{1}{2}
\right) , \label{model}
\end{eqnarray} 
where $c^{\dagger}_{i\alpha \sigma}$ ($c_{i\alpha \sigma}$) denotes an 
electron creation (annihilation) operator with orbital $\alpha$(=1,2) and
spin $\sigma$ at the $i$th site, and $n_{i\alpha \sigma}$ is 
$c^{\dagger}_{i\alpha \sigma} c_{i\alpha \sigma}$. ${\vec{S}}_{i\alpha 
}$ denotes electron spin operator with orbital $\alpha$ at the $i$th site. 
Concerning the hopping matrix elements, the nearest neighbor 
hopping between the same type of orbitals is assumed, 
$t^{\alpha \alpha '}_{i,i+1}=t \delta_{\alpha \alpha '}$. 
The simplest system which shows this property is illustrated 
in Fig.\,\ref{model_fig}: the $p_x$ and $p_y$ orbitals along 
a chain parallel to the $z$ axis. 
We are interested in the case where $t$, $U$, and $J$ are positive.

To study the region of strong correlation, we consider the limit of 
$U,J \gg t$ at quarter filling. In this case charge degrees 
of freedom are suppressed and 
the system becomes a Mott insulator. The effective Hamiltonian 
obtained by the usual second order perturbation is
\begin{eqnarray}
& &H_{{\rm{eff}}}=\sum_{i} \Bigg{\{} \frac{4t^{2}}{U}
\left({\vec{S}}_{i}\cdot {\vec{S}}_{i+1}-\frac{1}{4}\right)
\left(2{T}^{z}_{i}\cdot {T}^{z}_{i+1}+\frac{1}{2}\right) \nonumber\\
&+&\frac{4t^{2}}{U+J}
\left({\vec{S}}_{i}\cdot {\vec{S}}_{i+1}-\frac{1}{4}\right)
\left({\vec{T}}_{i}\cdot{\vec{T}}_{i+1}-2{T}^{z}_{i}
\cdot {T}^{z}_{i+1}+\frac{1}{4}\right) \nonumber \\
&+&\frac{4t^{2}}{U-J}
\left({\vec{S}}_{i}\cdot {\vec{S}}_{i+1}+\frac{3}{4}\right)
\left({\vec{T}}_{i}\cdot{\vec{T}}_{i+1}-\frac{1}{4}\right)
\Bigg{\}} \quad , \label{heff}
\end{eqnarray}
where
\begin{eqnarray*}
{\vec{S}}_{i} \equiv \frac{1}{2}\sum_{\alpha\sigma\sigma'}\left(
c^{\dagger}_{i\alpha\sigma}{\vec{\tau}}_{\sigma\sigma '}c_{i\alpha \sigma '}
\right)
\end{eqnarray*}
are the spin operators and 
\begin{eqnarray*}
{\vec{T}}_{i} \equiv \frac{1}{2}\sum_{\sigma\alpha\alpha'}\left(
c^{\dagger}_{i \alpha\sigma}{\vec{\tau}}_{\alpha \alpha '}c_{i\alpha '\sigma }
\right)
\end{eqnarray*}
are the pseudospin operators 
which describe the orbital degrees of freedom. 
In the above equations ${\vec{\tau}}$ are the Pauli spin matrixes.

As a first step, we consider the case with the highest symmetry by taking 
the $J\rightarrow 0$ limit. 
Then $H_{\rm{eff}}$ becomes, neglecting a constant term, 
\begin{equation}
H_{\rm{eff}}=K\sum_{i}P ^{( S={1}/{2})}_{i,i+1}
\cdot P ^{( T={1}/{2})} _{i,i+1} ,
\label{heff2}
\end{equation}
where $K\equiv 2t^{2}/U$, 
$P_{i,i+1}^{(S={1}/{2})}\equiv
2{\vec{S}}_{i}\cdot{\vec{S}}_{i+1}$+1/2, and 
$P_{i,i+1}^{(T={1}/{2})}\equiv
2{\vec{T}}_{i}\cdot{\vec{T}}_{i+1}$+1/2. 
$P_{i,i+1}^{(S={1}/{2})}$ and $P_{i,i+1}^{(T={1}/{2})}$
are the spin-1/2 and the pseudospin-1/2 exchange operators 
between the $i$th and ($i$+1)-th sites, respectively. 

Since the Hamiltonian (\ref{heff2}) exchanges both $S$ and $T$ spins 
at the same time,
the spin and orbital degrees of freedoms are 
combined into the SU(4) spin (denoted by ${\vec{S}}_{i}^{({3}/{2})}$) and 
$H_{\rm{eff}}$ is described 
by using spin-3/2 exchange operators as follows:
\begin{equation}
H_{\rm{SU(4)}}=K\sum_{i}P ^{( S={3}/{2})}_{i,i+1},
\label{hsu4}
\end{equation}
where 
\begin{eqnarray*}
P ^{( S={3}/{2})}_{i,i+1}=
& \frac{2}{9}  ({{\vec{S}}^{({3}/{2})}_{i}}
\cdot{{\vec{S}}^{({3}/{2})}_{i+1}})^3
+\frac{11}{18}({{\vec{S}}^{({3}/{2})}_{i}}
\cdot{{\vec{S}}^{({3}/{2})}_{i+1}})^2 \\
&-\frac{9}{8}  ({{\vec{S}}^{({3}/{2})}_{i}}
\cdot{{\vec{S}}^{({3}/{2})}_{i+1}})
-\frac{67}{32}.
\end{eqnarray*}
The Hamiltonian clearly has the SU(4) symmetry. 
We call this Hamiltonian the SU(4) exchange Hamiltonian.
 
The exact ground state energy and the dispersion relations 
of the SU(4) exchange Hamiltonian have been already obtained 
by the application of the Bethe ansatz technique to the 
higher spin-chain problems\cite{bethe}. 
In this paper, we investigate this model 
as the coupled spin and orbital system 
with the strongest orbital quantum fluctuations. 
It is worth noting that 
the SU(4) exchange Hamiltonian 
may play a similar role as the SU(2) 
AF Heisenberg model for the single orbital Hubbard model. 
The assumptions that the hoppings of electrons 
are possible only between the same orbitals 
and the vanishing $J$ produce this SU(4) symmetry, 
independently of the strength of the Coulomb repulsion $U$. 
Generally speaking, in real materials the Hund rule coupling $J$ 
is not small. However, an understanding of the most symmetric case 
will be important for future studies of less symmetric cases 
corresponding to a finite $J$.
\vspace*{-3ex}

\section{Ground state and excitations}
To understand the physics of the present model, let us consider
 $S$ and $T$ spins as classical spins. The spin configurations 
where every adjacent two, either $S$ or $T$, spins point the opposite 
direction have the lowest energy. Thus the classical 
ground state energy is 
zero and the degeneracy of the ground states is macroscopic in the 
classical theory. This situation is different from usual orbital and/or 
spin orderings discussed so far. 
Thus it is essential to examine the properties of Hamiltonian
(\ref{hsu4}) by unbiased methods.

\begin{figure}
\epsfxsize=86mm
\epsffile{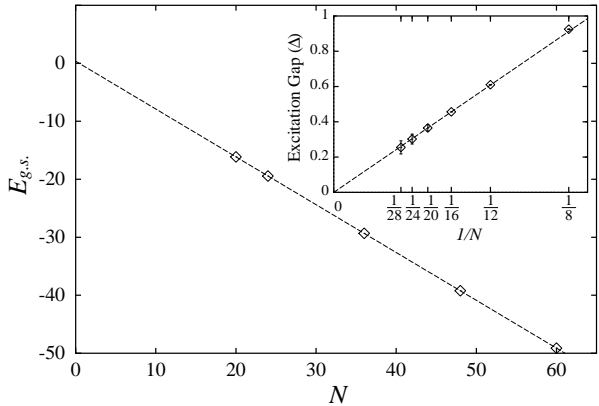}
\vspace*{2mm}
\caption{Ground state energy as a function of $N$. 
The numerical errors, which are
estimated from the truncation errors in the DMRG calculation, are less than 
0.01\%. The broken lines represent the linear fitting; 
$E_{g.s.}=-0.825\times N + 0.35(3)$. 
The inset shows the excitation gap ($\Delta$) as a function of $1/N$. 
Error bars are estimated from the truncation errors 
in the DMRG calculation. 
The dotted line represents the linear fitting, $\Delta = 7.3/N$}
\label{gs_gap}
\end{figure}

First, we calculate the ground state energy ($E_{g.s.}$) by the 
DMRG method\cite{dmrg}
 in the subspace of 
$(S_{tot}^{z},T_{tot}^{z})$=(0,0), where 
$S_{tot}^{z}\equiv \sum_{i}S_{i}^{z}$ and 
$T_{tot}^{z}\equiv \sum_{i}T_{i}^{z}$. 
We take $K=1$ as the energy unit 
and use here the open boundary conditions (OBC) 
to obtain sufficient accuracy by the DMRG method. 
The obtained results are shown in Fig.\,\ref{gs_gap}, which shows 
that the ground state energy per site ($E_{g.s.}/N$) 
and the surface energy are equal to $-$0.825 and 0.35(3), 
respectively. 
This ground state energy is, of course, consistent with that obtained 
by the Bethe ansatz, $E_{g.s.}/N \simeq -0.825\,12$~\cite{bethe}.

In Fig.\,\ref{gs_gap} the ground state energies are 
plotted only for $N=4n$, for which 
the minimum energy in the subspace of
$(S_{tot}^{z},T_{tot}^{z})$=(1,1) 
is different from 
$E_{g.s.}$. From Table I it is seen that the subspace 
$(S_{tot}^{z},T_{tot}^{z})$=(0,0) is included in every 
irreducible representation, 
but $(S_{tot}^{z},T_{tot}^{z})$=(1,1) belongs to 
any irreducible representation except 
for $[1^{4}]$. 
Thus it is concluded that 
the ground state belongs to the $[1^{4}]$ irreducible representation 
in the Young's diagram notation. 
Similarly, by calculating the ground state energies with 
changing $(S_{tot}^{z},T_{tot}^{z})$, 
it is found that for $N=4n+2$ the ground states belong to either
$[2^3]$ or $[2^{2}1^{2}]$, 
which are degenerate 10- and 6-fold, respectively (see Table I). 

These quantum numbers can be understood from the point of view 
of maximum antisymmetrization. That is, 
the irreducible representations thus 
obtained for the ground states are compatible with 
the simple fact that the more antisymmetric part one irreducible 
representation has, the lower is its ground state energy in the subspace, 
because the Hamiltonian (\ref{hsu4}) is 
the sum of SU(4) exchange operators. 
To avoid complications coming from the degenerate ground states, 
we consider the systems 
of $N=4n$ in the following. 
In this case the ground state belongs to the $[1^{4}]$ and is 
a singlet. Since the Lieb-Schultz-Mattis theorem applies to this model, 
the excitations are expected to be gapless, 
provided that no other symmetry is broken, 
in the same way as the spin-1/2 AF Heisenberg model 
with SU(2) symmetry.

To estimate the excitation energy, 
we calculate the ground state energy and the first excited state energy 
by using the DMRG method for $N \ge 20$ and by 
the exact diagonalization (Lanczos method) for $N \le 16$.
We determine the ground state energy and the first excited energy 
by the minimum energies of the states whose quantum numbers 
$(S_{tot}^{z},T_{tot}^{z})$ are specified: $(0,0)$, $(1,1)$, etc. 
It is found that the first excited state belongs to $[2^{1}1^{2}]$.

Though the DMRG is more suitable for OBC than periodic boundary 
conditions (PBC), here we apply the PBC in order to study the 
properties in the bulk limit. 
When we use the OBC, we get lower excitation energies than those shown
in the inset of Fig.\,\ref{gs_gap}, 
but they correspond to excitations at the surfaces
rather than those in bulk. 
From the inset of Fig.\,\ref{gs_gap} we can conclude 
that the excitation gap $(\Delta)$ 
goes to zero as $\Delta \sim 7.3/N$.

\begin{figure}
\epsfxsize=86mm
\epsffile{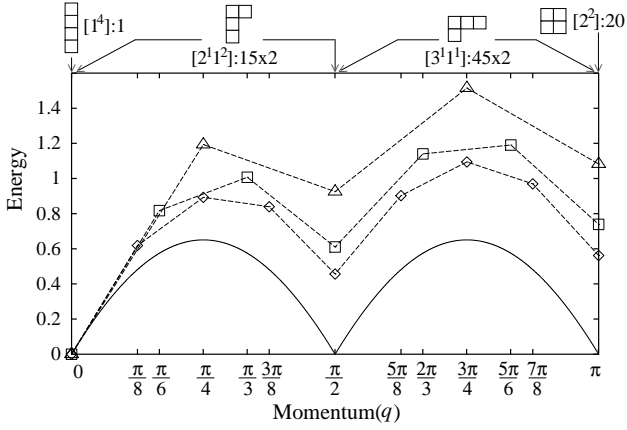}
\vspace*{2mm}
\caption{Dispersion relations. 
The symbols $\triangle$, $\Box$, and $\Diamond$ 
represent data for $N=8$, 12, and 16, respectively. 
Young's diagrams (YD) show the irreducible representations. 
The numbers accompanied by YD 
show the degeneracy and $\times 2$ represents 
the same weight from $q$ and $-q$. 
The solid line represents the Bethe ansatz result[7].
}
\label{disp}
\end{figure}

In order to examine the properties of the excitations in more detail, 
we calculate the dispersion relation by using the Lanczos method 
with the use of translational symmetry for the systems with the PBC. 
Figure \ref{disp} shows that the excitation spectrum 
has a ``bactrian camel'' structure and shows softening at $q=\pi/2$. 
This structure is also known by the Bethe ansatz results~\cite{bethe}. 
Corresponding to the softening at $q=\pi/2$, the correlation functions 
would show a characteristic feature, 
namely, oscillatory behaviors with a period of four, 
which we will discuss in the next section. 

Figure \ref{disp} shows that height of the left hump is always 
lower than that of the right one. 
To consider a possible reason, 
we determine the irreducible representation of each state for 
$N=8$ and $12$. Quantum numbers assigned for each $q$ point in 
the dispersion curves are shown in Fig.\,\ref{disp} by the SU(4) 
Young's diagram representations. 
In these finite size calculations, 
the state at $q=\pi/2$ and left part of the two humps always belongs 
to $[2^{1}1^{2}]$ and the right one to $[3^{1}1^{1}]$. 
From Table I, the first excited states at $q=\pi/2$ 
consist of the coupled spin and orbital excitations 
in addition to the pure spin and orbital excitations and 
have the $15\!\times\!2$-fold degeneracy in total. 
The difference of the height of the two humps may be attributed to 
the difference of the irreducible representations of the two 
humps. In fact, the left (lower) hump belongs to 
the irreducible representation which has a more antisymmetric part 
than that of the right (higher) one. 
In the bulk limit, however, we expect that the two parts, 
$0<q<\pi/2\,$ and $\,\pi/2<q<\pi$, converge to the same dispersion 
relation as is known by the Bethe ansatz solution\cite{bethe}.

\begin{table}
\caption{Irreducible representations of the SU(4) symmetry 
given by Young's diagrams (YD) and their relations to 
$(S_{tot},T_{tot})$ and $S^{(3/2)}_{tot}$ representations[9], 
where $S_{tot}$, $T_{tot}$, and $S^{(3/2)}_{tot}$ are the 
magnitudes of the 
${\vec{S}}_{tot}\equiv \sum_{i}{\vec{S}}_{i}$, 
${\vec{T}}_{tot}\equiv \sum_{i}{\vec{T}}_{i}$, and 
${\vec{S}}^{(3/2)}_{tot}\equiv \sum_{i}{\vec{S}}^{(3/2)}_{i}$, 
respectively. 
Here $\nu$ is the degeneracy of each representations.}
\label{YD}
\begin{tabular}{cccc}
YD&$\nu$&$(S_{tot},T_{tot})$&$S^{(3/2)}_{tot}$ \\ \hline
$[1^{4}]$     & 1&(0,0)&0 \\
$[2^{1}1^{2}]$&15&(0,1)$\oplus$(1,0)$\oplus$(1,1) & 1$\oplus$2$\oplus$3 \\
$[2^{2}]$&20&(0,0)$\oplus$(1,1)$\oplus$(0,2)$\oplus$(2,0) & 
0$\oplus$2$\oplus$2$\oplus$4 \\
$[3^{1}1^{1}]$&45&(0,1)$\oplus$(1,0)$\oplus$(1,1)$\oplus$(1,2)$\oplus$(2,1) &
1$\oplus$1$\oplus$2$\oplus$3$\oplus$3$\oplus$4$\oplus$5 \\
$[4^{1}]$     &35&(0,0)$\oplus$(1,1)$\oplus$(2,2) & 
0$\oplus$2$\oplus$3$\oplus$4$\oplus$6 \\ \hline
$[2^{2}1^{2}]$& 6&(1,0)$\oplus$(0,1)& 0$\oplus$2 \\
$[2^{3}]$     &10&(0,0)$\oplus$(1,1)& 1$\oplus$3
\end{tabular}
\end{table}

\section{Correlation Functions}
Now we move on to the behaviors of the correlation functions, 
$\langle {S}^{z}_{i}\cdot {S}^{z}_{i+j}\rangle _{g.s.}$, 
where $\langle \cdots \rangle _{g.s.}$ denotes expectation values 
for the ground state. Since Hamiltonian (\ref{heff2}) has rotational 
symmetry with respect to both $S$ and $T$ spins, we consider only 
$z$ components of spins.

We use the OBC to get better accuracy in the DMRG calculations, 
but in this case we must keep in mind that the data contain the 
effects from boundaries. 
$\langle S^{z}_{i} \cdot S^{z}_{i+j}\rangle_{g.s.}$ shows an 
oscillatory behavior with a period of four as a function of $j$. 
But the correlation functions also vary with a period of four with 
respect to $i$. That is, $\langle S^{z}_{i} \cdot S^{z}_{i+j}\rangle 
_{g.s.}$ is equal to $\langle S^{z}_{i+4} \cdot S^{z}_{i+4+j}\rangle 
_{g.s.}$ for any $i,j$.

\begin{figure}
\epsfxsize=86mm
\epsffile{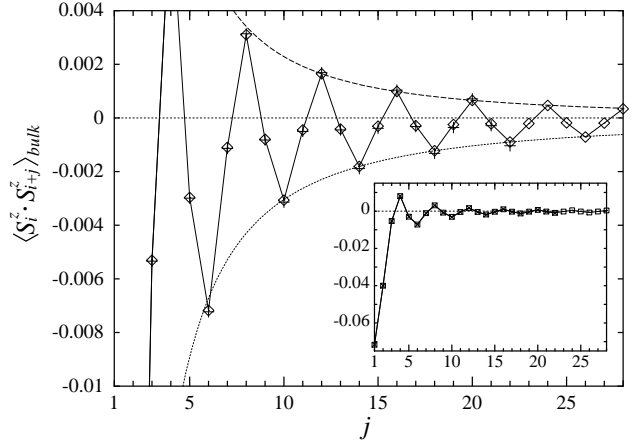}
\vspace*{2mm}
\caption{Correlation functions ($\langle S^{z}_{i} 
\cdot S^{z}_{i+j}\rangle _{bulk}$) for the systems with 
$N=48$ ($+$) and $60$ ($\Diamond$). 
Broken and dotted lines are $0.144/j^{1.80}$ and
$-0.108/j^{1.55}$, which are the results of the least mean square
fitting using the upper and lower data, respectively. Numerical
errors, which are estimated from the values of $\langle S^{z}_{i}\cdot
T^{z}_{i+j}\rangle _{bulk}$, are less than 1\%.
The inset shows the entire form of the correlation functions.
The symbols $\times$ and $\Box$ represent data 
for $N=48$ and $60$, respectively, but they overlap nearly perfectly.}
\label{corr}
\end{figure}

This behavior is caused by the standing wave with a period of four
originating from the open boundaries. To remove such an artifact due
to the OBC, we average $\langle S^{z}_{i} \cdot S^{z}_{i+j}\rangle
_{g.s.}$ for one period with respect to $i$. Thus we define an
approximate bulk correlation function as follows: 
\begin{equation}
\langle S^{z}_{i} \cdot S^{z}_{i+j}\rangle _{bulk} \equiv
\frac{1}{4} \sum_{k=0}^{3} \langle {S}^{z}_{i+k}\cdot {S}^{z}_{i+k+j} 
\rangle_{g.s.} \quad. \label{bulkcorr}
\end{equation}
After this averaging procedure, we get the natural behavior of the 
correlation functions as shown in Fig.\,\ref{corr} .

Because the results discussed in the previous section show 
that this model is gapless, we try to fit
the envelope of $\langle S^{z}_{i} \cdot S^{z}_{i+j}\rangle _{bulk}$
data with a power-law function ( $j^{-\alpha}$) by the least mean
square method and get critical exponent $\alpha$ equal to 1.80 or 
1.55 depending on using either the upper ($j=12,$ 16, and 20) 
or the lower ($j=14$, 18, and 22) data. 
We did not use the data of $j=24, 26$ and $28$,
because these sites are too close to the boundary. Due to finite size effects,
the value of $\alpha$ depends on how to fit, but $\alpha$ is always
between 1.5 and 2.0.
From these results, we conclude that the asymptotic form of the
correlation function is given by
\begin{equation}
\langle {S}^{z}_{i}\cdot {S}^{z}_{i+j} \rangle _{bulk}
\sim \frac{ \cos{ (\frac{\pi}{2}j)}}
{j^{\alpha}} ;\quad \alpha = 1.5 - 2.0 , \label{asympt}
\end{equation}
in the bulk limit.

Of course, $\langle S^{z}_{i} \cdot S^{z}_{i+j}\rangle _{bulk}$ is
equal to $\langle T^{z}_{i} \cdot T^{z}_{i+j}\rangle _{bulk}$, because
of the symmetry of Hamiltonian (\ref{heff2}) 
concerning the exchange between ${{\vec{S}}}$ and
${{\vec{T}}}$. Furthermore, $\langle S^{z}_{i} \cdot T^{z}_{i+j}\rangle
_{bulk}$ always equals zero as is easily shown by the Wigner-Eckart
theorem. In fact, the calculated values of $\langle S^{z}_{i} \cdot
T^{z}_{i+j}\rangle _{bulk}$ are almost zero, and the numerical errors
for the values of $\langle S^{z}_{i} \cdot S^{z}_{i+j}\rangle _{bulk}$
may be estimated from these values, which are less than $1\%$ even at
the farthest site from $i$.

\begin{figure}
\epsfxsize=86mm
\epsffile{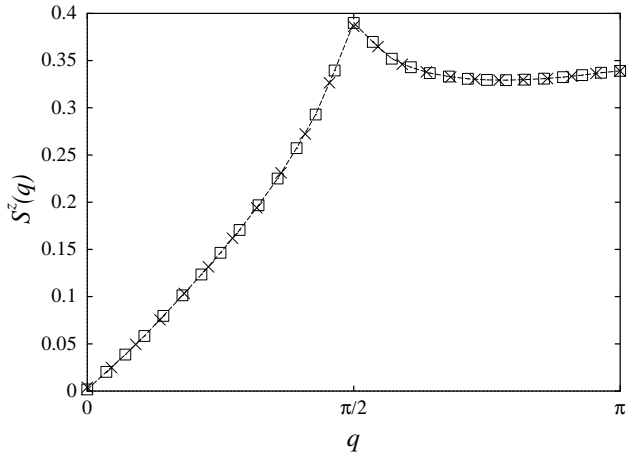}
\vspace*{2mm}
\caption{Fourier transformation of the 
correlation function 
for the systems with $N=60$ ($\Box$) and $N=48$ ($\times$).}
\label{sq}
\end{figure}

Next we study the structure factor defined by 
\begin{equation}
S^{z}(q) \equiv \sum_{j=-N/2+3}^{N/2-2}\langle S^{z}_{i}\cdot
S^{z}_{i+j} \rangle_{bulk} \cdot e^{-iqj} \quad . \label{sq_eq}
\end{equation}
As is seen in Fig.\,\ref{sq}, $S^{z}(q)$ has a characteristic cusp
structure at $q=\pi/2$. This result is consistent with the softening
at $q=\pi/2$ in the dispersion relation.
By Fourier transformation of Eq.\,(\ref{asympt}), the
analytic form of $S^{z}(q)$ is given by
\begin{equation}
S^{z}(q) \sim S^{z}(q=\frac{\pi}{2})-\frac{\pi}{2}
\frac{|q-\frac{\pi}{2}|^{\alpha-1}}
{{\it{\Gamma}}(\alpha)\sin{\frac{\alpha-1}{2}\pi}}
+O\left(q-\frac{\pi}{2}\right) \label{anasq}
\end{equation}
around $q=\pi/2$, when $1<\alpha <3$. 
If $\alpha$ is greater than 2, Eq.\,(\ref{anasq}) does not
show any cusp structure at $q=\pi/2$. So $\alpha$ must be less than
$2$, since we clearly see 
the cusp structure of $S^{z}(q)$ which becomes sharper as the 
system size is increased. 

By the SU(4) conformal field theory, 
the critical exponent of the SU(4) spin correlation functions 
with $q=\pi/2$ oscillations 
is obtained to be 3/2\cite{critical}, 
which is consistent with the present numerical result. 
Although the correlation functions discussed 
in this paper are the $S$-spin correlation functions 
but not the SU(4) spin correlation functions, 
we can show that the exponent is the same for 
the two correlation functions. 

\section{Conclusions and discussions}
In conclusion, we have studied the quantum critical state for the 
coupled spin-orbit system. 
The quantum numbers of the ground state and the lowest branch of the 
excitations are determined. 
Furthermore, the spin-spin correlation 
functions are obtained explicitly for the first time by the 
DMRG method. It shows a 
power-law decay with a period of four, 
which originate from the interference between 
the spin and orbital degrees of freedom. 
The exponent of the asymptotic behavior is consistent with the 
prediction by the conformal field theory. 

In this paper we have investigated only the most symmetric model,
but it is more realistic to consider a model with lower symmetry
corresponding to a finite $J$.
In such a case the effective Hamiltonian is given by 
the $S$-spin isotropic and $T$-spin Ising-type 
Hamiltonian (\ref{heff}), whose properties are not yet fully understood. 

Related to the SU(4) model, several models with lower symmetries 
have been studied\cite{shibata,kawano,mikeska}. 
Kawano and Takahashi\cite{kawano} discussed $S$-spin isotropic and $T$-spin
$XY$-type Hamiltonians to study the three-leg antiferromagnetic Heisenberg
ladder and showed that such a model is gapfull 
and has exponentially decaying
correlation functions. Kolezhuk and Mikeska\cite{mikeska} studied a 
special SU(2)$\times$SU(2) symmetric Hamiltonian $\sum_i
({\vec{S}}_{i}\cdot{\vec{S}}_{i+1}+3/4)({\vec{T}}_{i}\cdot{\vec{T}}_{i+1}+3/4)$ 
and showed that this model is also gapfull.

It may be possible to study the properties of these models in 
a unified way by introducing different type of  
anisotropies from the SU(4) symmetric point. For this purpose, 
it is highly desirable to develop an analytic theory 
around this symmetric point.

\acknowledgments We would like to thank Norio Kawakami for pointing out 
the exact results of the SU(4) model and many helpful comments. 
Thanks are also due to Manfred Sigrist for valuable discussions. 
We are grateful to Fu Chun Zhang who sent us his related work 
prior to publication\cite{FuChun}. 
The numerical exact diagonalization calculations of this work were 
done by using TITPACK ver.\,2 developed by H. Nishimori. 
This work is financially supported by a Grant-in-Aid for Scientific 
Research on Priority Areas from the Ministry of Education, Science,
Sports and Culture. N.S. is supported by the Japan Society for the 
Promotion of Science.

\end{multicols}


\begin{references}
\bibitem[*]{address}Present address: 
Institute of Applied Physics, University of Tsukuba, 
Tsukuba 305, Japan.
\bibitem{kugel} K. I. Kugel and D. I. Khomskii, 
Sov. Phys. JETP {\bf 37}, 725 (1973)
[Zh. Eksp. Teor. Fiz. {\bf 64}, 1429 (1973)]. 
\bibitem{inagaki} S. Inagaki, 
J. Phys. Soc. Jpn. {\bf 39}, 596 (1975). 
\bibitem{shiinaC} R. Shiina, H. Shiba, and P. Thalmeier, 
J. Phys. Soc. Jpn. {\bf 66}, 1741 (1997). 
\bibitem{pearson} J. des Cloizeaux and J. J. Pearson, 
Phys. Rev. {\bf 128}, 2131 (1962). 
\bibitem{lieb} E. H. Lieb, T. Schultz, and D. J. Mattis, 
Ann. Phys. (N.Y.) {\bf 16}, 407 (1961).
\bibitem{affleck} I. Affleck and E. H. Lieb, 
Lett. Math. Phys. {\bf 12}, 57 (1986).
\bibitem{bethe} B. Sutherland, 
Phys. Rev. B {\bf 12}, 3795 (1975). 
\bibitem{dmrg} S. R.  White, 
Phys. Rev. Lett. {\bf 69}, 2863 (1992). 
\bibitem{group} X. Hamermesh, {\it Group Theory}, 
(Addison-Wesley, Reading, MA, 1962). 
\bibitem{critical} I. Affleck, 
Nucl. Phys. B {\bf 265}, 409 (1986).
\bibitem{shibata} N. Shibata, M. Sigrist, and E. Heeb, 
Phys. Rev. B {\bf 56}, 11084 (1997). 
\bibitem{kawano} K. Kawano and M. Takahashi, 
J. Phys. Soc. Jpn. {\bf 66}, 4001 (1997). 
\bibitem{mikeska} A. K. Kolezhuk and H.-J. Mikeska, 
Phys. Rev. Lett. {\bf 80}, 2709 (1998).
\bibitem{FuChun} Y. Q. Li, Michael Ma, D. N. Shi, and F. C. Zhang, 
cond-mat/9804157 .
\end{references}
\end{document}